\documentclass[twocolumn,prl]{revtex4}
\usepackage[T1]{fontenc}
\usepackage{amsmath,amsbsy,amssymb,graphicx}
\usepackage{times}

\def\beginABC{\begin{subequations}}
\def\endABC{\end{subequations}}
\let\mathbf=\boldsymbol

\begin{document}

\title{{\Large Hexagonally Warped Dirac Cones and Topological Phase
Transition}\\
{\Large \ in Silicene Superstructure}}
\author{Motohiko Ezawa}
\affiliation{Department of Applied Physics, University of Tokyo, Hongo 7-3-1, 113-8656,
Japan }


\begin{abstract}
Silicene is a monolayer of silicon atoms forming a two-dimensional honeycomb
lattice. We investigate the topological properties of a silicene
superstructure generated by an external periodic potential. The
superstructure is a quantum spin-Hall (QSH) insulator if it is topologically
connected to silicene. It is remarkable that two inequivalent K and K'
points in the silicene Brillouin zone are identified in certain
superstructures. In such a case two Dirac cones coexist at the same Dirac
point in the momentum space and they are hexagonally warped by the Coulomb
interaction. We carry out a numerical analysis by taking an instance of the ($3\times 3$) superstructure on the ($4\times 4$) structure of the Ag
substrate. We show that it is a QSH insulator, that there exists no
topological phase transition by external electric field, and that the
hexagonally warping occurs in the band structure.
\end{abstract}

\maketitle


\section{Introduction}

Topological insulator is a new state of matter which is characterized by an
insulating bulk and surface edge modes\cite{Hasan,Qi}. It is robust against
perturbations such as disorders and impurities as far as the gap does not
close. It is an interesting and important question whether the topological
insulator is robust against a formation of a superstructure in a periodic
potential. This is nontrivial since it changes a global structure of the
material such as the Brilloin zone. It is intriguing that superstructures
have been realized in silicene on top of the Ag substrate\cite{GLayPRL,Kawai,Takamura}. There may well be other ways to create periodic
potentials. Silicene is an interesting play ground where many types of
superstructures are naturally materialized.

Silicene is a sheet of silicon atoms replacing carbons in graphene. It could
follow the trend of graphene and attract much attention\cite%
{GLayPRL,Kawai,Takamura,Shiraishi,Guzman,LiuPRL,LiuPRB,EzawaNJP,EzawaAQHE,EzawaEPJB,EzawaPhoto,EzawaJ}. It would open new perspectives for applications, especially due to its
compatibility with Si-based electronics. Almost every striking property of
graphene could be transferred to this innovative material. Indeed, it has
Dirac cones akin to graphene. It has additionally a salient feature, that is
a relatively large spin-orbit (SO) gap, which provides a mass to Dirac
electrons and realizes a detectable quantum spin Hall (QSH) effect\cite%
{LiuPRL,EzawaNJP}. The QSH insulator is a two-dimensional topological
insulator with helical gapless edge modes\cite{Hasan,Qi}. Furthermore a
topological phase transition occurs from the QSH insulator to the trivial
band insulator in the electric field\cite{EzawaNJP}. In this paper we
address the problem if a superstructure is a QSH\ insulator as well and if
there exists a similar topological phase transition.

In silicene the states near the Fermi energy are $\pi $ orbitals residing
near the inequivalent K and K' points at opposite corners of the hexagonal
Brillouin zone. It is folded into a reduced Brillouin zone in a
superstructure. We can generally argue that a superstructure is a QSH
insulator provided it is topologically connected to silicene, namely, the
gap is open and does not close as the periodic potential is continuously
switched off. We also study the problem whether a silicene superstructure
undergoes a topological phase transition by applying external electric field
as in free-standing silicene\cite{EzawaNJP}. Furthermore, we point out an
intriguing possibility that the K and K' points are identified by the
folding. When it happens, there exists only one Dirac point where two Dirac
cones coexist, and this is experimentally detectable. In addition to general
arguments, to demonstrate these novel phenomena we carry out a numerical
analysis by taking an instance of the ($3\times 3$) superstructure on the ($%
4\times 4$) structure of the Ag substrate\cite{NoteA}.

This paper is composed as follows. In Section II, we present how a
superstructure is constructed from a honeycomb lattice. We derive the
condition that the K and K' points coincide. The (3$\times $3)
superstructure is the simplest example where this coincidence occurs. In
Section III, we postulate the tight-binding Hamiltonian to describe a
superstructure. Making a numerical analysis of the band structure of the (3$%
\times $3) superstructure, we explicitly see that two degenerate Dirac cones
are merged into two undegenerate hexagonally warped cones in the presence of
infinitesimal Coulomb interactions. The mechanism of the hexagonal warping
is explained based on the effective low-energy Dirac theory. In Section IV
we investigate the topological properties of superstructures in three
different ways, i.e., the topological connectedness, the bulk-edge
correspondence and the Dirac theory. We present a general criterion when the
topological phase transition may occur in superstructures. In the instance
of the (3$\times $3) superstructure, it is shown to be topologically
connected to silicene, and hence it is a QSH insulator. We also show that
there exists no topological phase transition in external electric field.
Section V is devoted to discussions.

\section{Superstructure}

\begin{figure}[t]
\centerline{\includegraphics[width=0.32\textwidth]{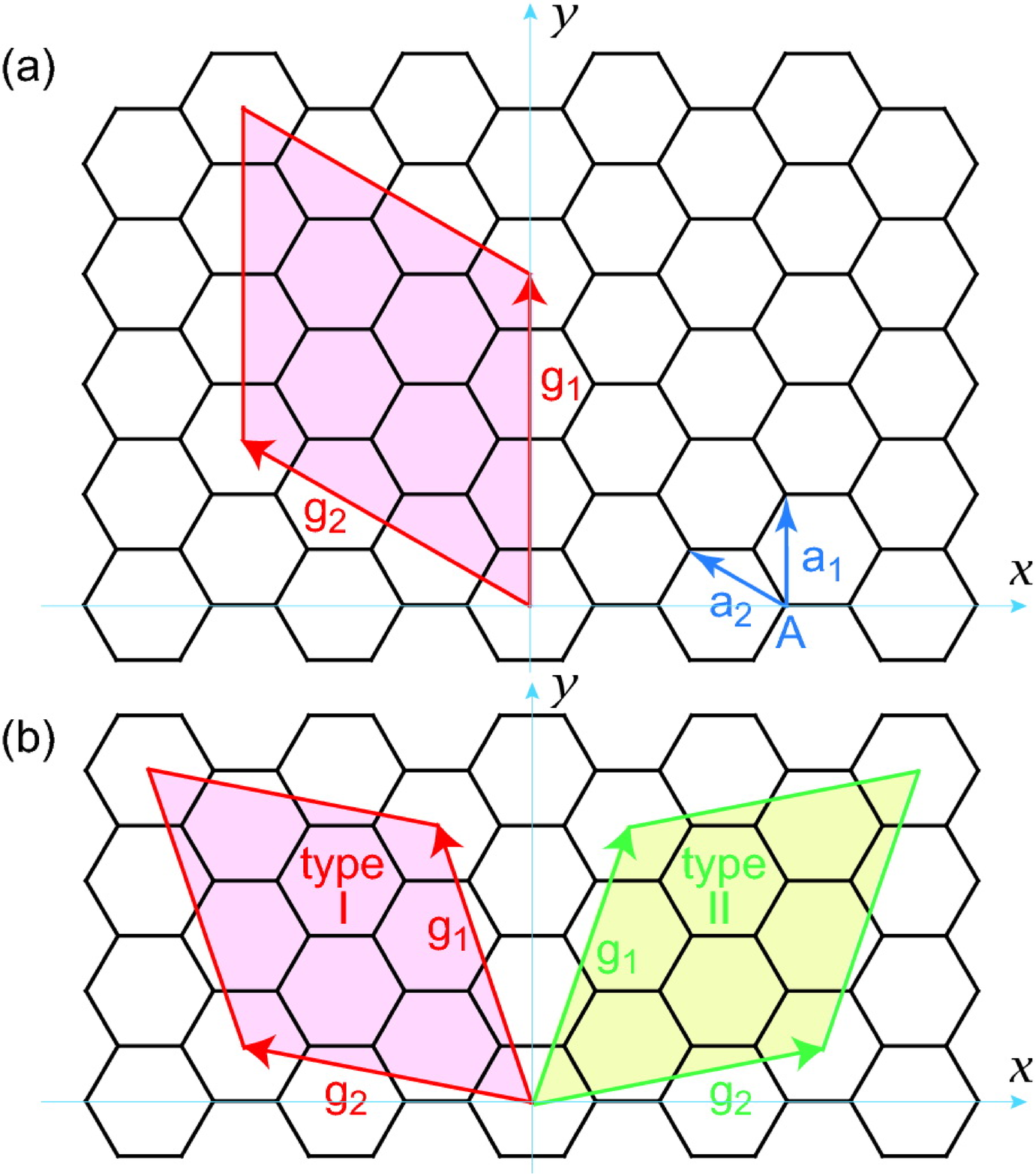}}
\caption{(Color online) Illustration of (a) ($3\times 3$) superstucture and
(b) ($\protect\sqrt{7}\times \protect\sqrt{7}$) superstructure. The shaded
area spanned by vectors $\boldsymbol{g}_{1}$ and $\boldsymbol{g}_{2}$
represents one unit cell of a superstructure.}
\label{FigSuperStruc}
\end{figure}

We start with a study of a superstructure made on the honeycomb lattice (Fig.%
\ref{FigSuperStruc}). The honeycomb lattice is specified by the two basis
vectors $\mathbf{a}_{1}$ and $\mathbf{a}_{2}$. It has two inequivalent sites
(A and B sites) per unit cell. There are three B sites adjacent to one A
site. The A and B sites are generated by\beginABC%
\begin{align}
\boldsymbol{A}(n_{1},n_{2})=& n_{1}\mathbf{a}_{1}+n_{2}\mathbf{a}_{2}, \\
\boldsymbol{B}(n_{1},n_{2})=& n_{1}\mathbf{a}_{1}+n_{2}\mathbf{a}_{2}+%
\mathbf{r}_{1},
\end{align}%
\endABC
where $\mathbf{r}_{1}$ is a vector connecting adjacent A and B sites, and $%
n_{i}$ are integers. They span the the A-sublattice and the B-sublattice. A
superstructure is specified by two translational vectors $\mathbf{g}_{1}$
and $\mathbf{g}_{2}$\ defined by%
\begin{equation}
\mathbf{g}_{1}=p\mathbf{a}_{1}+q\mathbf{a}_{2},\qquad \mathbf{g}_{2}=e^{\pm
i\pi /3}\mathbf{g}_{1},
\end{equation}%
with $p$ and $q$ integers subject to $0<p$. The choice of the angles $\pm
\pi /3$ generates the same superstructure due to the hexagonal symmetry.
There are two types: The type-I superstructure is generated by choosing $%
0\leq q\leq p$, while the type-II superstructure is generated by the space
inversion of the type-I. When $q=0$, type-I and type-II are identical. We
show two examples by choosing $(p,q)=(3,0)$ and $(2,1)$ in Fig.\ref%
{FigSuperStruc}.

By a geometrical analysis of the honeycomb lattice we have 
\begin{equation}
\left\vert \mathbf{g}_{1}\right\vert =\sqrt{p^{2}+q^{2}+pq}.
\end{equation}%
The structure thus generated is customarily referred to as the ($\left\vert 
\mathbf{g}_{1}\right\vert \times \left\vert \mathbf{g}_{1}\right\vert $)
superstructure. The number of silicon atoms per unit cell is given by%
\begin{equation}
N=2\left\vert \mathbf{g}_{1}\right\vert ^{2}=2\left( p^{2}+q^{2}+pq\right) .
\label{NumbeSI}
\end{equation}%
The angle $\phi $ between the vector $\mathbf{g}_{1}$ and the $y$-axis is
given by 
\begin{equation}
\tan \phi =\sqrt{3}q/(2p+q).
\end{equation}%
For instance, the choices $(p,q)=(3,0)$ and $(2,1)$ generate the ($3\times 3$%
) superstructure with $N=18$ and $\phi =0$ and the ($\sqrt{7}\times \sqrt{7}$%
) superstructure with $N=14$ and $\phi =\arctan (\sqrt{3}/5)$, respectively,
as illustrated in Fig.\ref{FigSuperStruc}.

\begin{figure}[t]
\centerline{\includegraphics[width=0.32\textwidth]{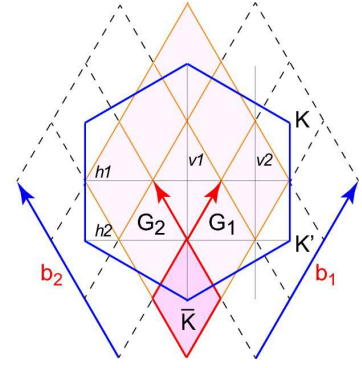}}
\caption{(Color online) Brillouin zone of silicene and the $(3\times 3$)
superstructure. Its shape is honeycom (blue lines) or equivalently rhombus
(large orange lines) for silicene, and rhombus (red lines) for the
superstructure. It is constructed by folding the Brillouin zone of silicene.
The K and K' points are identified, which we denote by \={K} and take at the
center of the Brillouin zone (the shaded area). The band structure along the
horizontal thin lines (\textit{h1,h2}) and vertical thin lines (\textit{v1,v2%
}) are shown in Fig.\protect\ref{FigS2Sa}.}
\label{FigSBril}
\end{figure}

The basis vectors $\mathbf{b}_{i}$ of the reciprocal lattice are given by
solving the relations $\mathbf{b}_{i}\cdot \mathbf{a}_{j}=2\pi \delta _{ij}$
in the honeycomb lattice (Fig.\ref{FigSBril}). The Brillouin zone is made by 
$\mathbf{b}_{1}$ and $\mathbf{b}_{2}$ in the reciprocal lattice with
opposite sides being identified. A similar construction is carried out for a
superstructure. The reciprocal vectors $\mathbf{G}_{1}$ and $\mathbf{G}_{2}$%
\ of a superstructure is given by 
\begin{equation}
\mathbf{G}_{i}=(\mathbf{g}_{i}\times \mathbf{n)/}\left\vert \mathbf{g}%
_{1}\times \mathbf{g}_{2}\right\vert ,  \label{RecipG}
\end{equation}%
where $\mathbf{n}$ is the unit vector perpendicular to the silicene plane.
The Brillouin zone of a superstructure is constructed from that of silicene
by identifying two momentum vectors $\mathbf{k}$ and $\mathbf{k}^{\prime }$
provided they are different only by the principal reciprocal vectors $%
\mathbf{G}_{1}$ and $\mathbf{G}_{2}$, as illustrated in Fig.\ref{FigSBril}.

There may occur an intriguing phenomenon. In general, when there are two
integers $n_{i}$ such that 
\begin{equation}
\mathbf{K}-\mathbf{K}^{\prime }=n_{1}\mathbf{G}_{1}+n_{2}\mathbf{G}_{2},
\label{CondiOneK}
\end{equation}%
the K and K' points are identified. We call it the \={K} point, and choose $%
\mathbf{\bar{K}}=(0,0)$. Two Dirac cones coexist at the \={K} point. This
happens indeed in the $(3\times 3)$ superstructure, where the Brillouin zone
is obtained by folding that of silicene three times along the $k_{x}$-axis
and three times along the $k_{y}$-axis. The area of the Brillouin zone is $%
1/9$ of that of silicene, and its shape is rhombus (Fig.\ref{FigSBril}).
This superstructure is simplest and most interesting, which we investigate
as an explicit example.

\section{Tight-Binding Hamiltonian}

The superstructure system is described by the second-nearest-neighbor
tight-binding model constructed as follows. The Hamiltonian consists of two
parts. The basic part is the Hamiltonian of silicene,\cite{KaneMele}

\begin{equation}
H_{0}=-t\sum_{\left\langle i,j\right\rangle \alpha }c_{i\alpha }^{\dagger
}c_{j\alpha }+i\frac{\lambda _{\text{SO}}}{3\sqrt{3}}\sum_{\left\langle
\!\left\langle i,j\right\rangle \!\right\rangle \alpha \beta }\nu
_{ij}c_{i\alpha }^{\dagger }\sigma _{\alpha \beta }^{z}c_{j\beta },
\label{BasicHamil}
\end{equation}%
where $c_{i\alpha }^{\dagger }$ creates an electron with spin polarization $%
\alpha $ at site $i$, and $\left\langle i,j\right\rangle /\left\langle
\!\left\langle i,j\right\rangle \!\right\rangle $ run over all the
nearest/next-nearest neighbor hopping sites. The first term represents the
usual nearest-neighbor hopping with the transfer energy $t=1.6$eV. The
second term represents the intrinsic SO coupling with $\lambda _{\text{SO}%
}=3.9$meV, where $\mathbf{\sigma }=(\sigma _{x},\sigma _{y},\sigma _{z})$ is
the Pauli matrix of spin, $\nu _{ij}=\left( \mathbf{d}_{i}\times \mathbf{d}%
_{j}\right) /\left\vert \mathbf{d}_{i}\times \mathbf{d}_{j}\right\vert $
with $\mathbf{d}_{i}$ and $\mathbf{d}_{j}$ the two bonds connecting the
next-nearest neighbors. Additionally there exist two types of Rashba SO
couplings ($\lambda _{\text{R1}}$ and $\lambda _{\text{R2}}$) to describe
silicene\cite{EzawaAQHE}. They are quite small, and play no important roles
in the present analysis, as far as we have numerically checked. To avoid
unnecessary complications, we neglect them.

\begin{figure}[t]
\centerline{\includegraphics[width=0.45\textwidth]{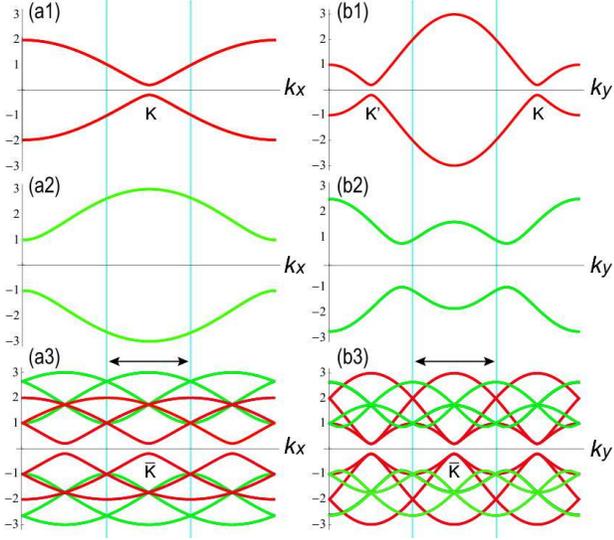}}
\caption{(Color online) (a1,a2) Band structure of silicene along the
horizontal lines (\textit{h1,h2}) in Fig.\protect\ref{FigSBril}. (b1,b2)
Band structure of silicene along the vertical lines (\textit{v1,v2}) in Fig.%
\protect\ref{FigSBril}. (a3,b3) 3-folded band structure obtained from
(a1,a2) together with shifts along the $k_{x}$axis, and the one obtained
from (b1,b2) together with shifts along the $k_{y}$-axis. They give the band
structure of the ($3\times 3$) superstructure shown in Fig.\protect\ref%
{FigBand}(a1) and (b1) with $U^{\prime }\rightarrow 0$. The horizontal arrow
in (a3,b3) indicates the first Brillouin zone.}
\label{FigS2Sa}
\end{figure}

The second part is the periodic potential term $H_{1}$ that introduces a
superstructure to the honeycomb lattice. The periodicity is such that the
band structure satisfies%
\begin{equation}
\varepsilon (\mathbf{k}+\mathbf{G}_{i})=\varepsilon (\mathbf{k}),
\label{BrillIdent}
\end{equation}%
where $\mathbf{G}_{i}$ are the principal reciprocal vectors (\ref{RecipG}).
It is to be emphasized that the band structure is simply constructed by
folding that of silicene when $H_{1}\simeq 0$, as illustrated in Fig.\ref%
{FigS2Sa} for an instance of the $(3\times 3)$ superstructure.

\subsection{An Explicit Example and Numerical Results}

\begin{figure}[t]
\centerline{\includegraphics[width=0.35\textwidth]{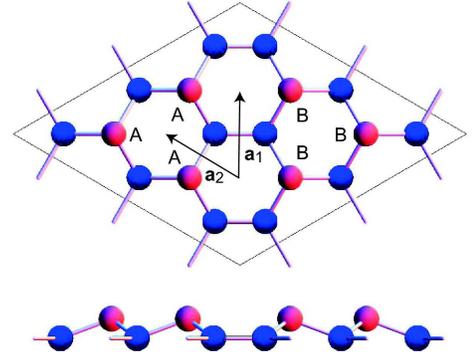}}
\caption{(Color online) Unit cell of the (3$\times $3) superstructure. There
are $18$ silicon atoms per unit cell, in which $6$ atoms (red) are in higher
position and $12$ atoms (blue) are in lower position. There are $3$ A and $3$%
\ B sites in higher position. Their vertical sepation is $2\ell \approx 0.8$%
\AA . The basis vectors are represented by $\mathbf{a}_{i}$.}
\label{FigUnit}
\end{figure}

Superstructures have been materialized on the Ag substrate. The structure of
the substrate acts as a periodic potential to silicene and generate a
superstructure on silicene. Such an effect can be formulated into the
periodic potential term $H_{1}$.

Silicene available now is mostly grown on the Ag(111) substrate. The most
common one is the ($3\times 3$) superstructure synthesized on the ($4\times
4 $) Ag-structure\cite{GLayPRL,Kawai}, though many other superstructures are
possible due to the commensurability with the Ag substrate\cite%
{Jamgo,Enriquez}. There are $3\times 3$ silicon atoms on top of $4\times 4$
silver atoms in this structure. The unit cell contains $2\times (3\times
3)=18$ silicon atoms. The superstructure contain two sublattices: There are $%
6$ ($12$) atoms in the higher (lower) sublattice which is $3.0$\AA\ ($2.2$%
\AA ) above the Ag surface, as illustrated in Fig.\ref{FigUnit}. This
structure has been observed in scanning tunneling microscopy (STM) images,
where only the atoms in the higher sublattice are visible\cite{GLayPRL,Kawai}%
. We may summarize the effect from the Ag substrate as a chemical potential
difference $U$ between silicon atoms in higher and lower sublattices.
Additionally we may apply the electric field $E_{z}$ perpendicular to the
plane. It generates a staggered sublattice potential $\varpropto 2\ell E_{z}$
with $2\ell \approx 0.8$\AA\ between them.

The basic nature of the above superstructure is summarized into the periodic
potential term, 
\begin{equation}
H_{1}=U^{\prime }\sum_{i\in \text{high}}c_{i}^{\dagger }c_{i},
\label{PotenU}
\end{equation}%
with%
\begin{equation}
U^{\prime }=U+2\ell E_{z},
\end{equation}%
where the summation is taken over higher sites: $U$ is a constant of the
order of $0.1$eV\ according to the first-principle calculation\cite{Enriquez}%
.

\begin{figure}[t]
\centerline{\includegraphics[width=0.47\textwidth]{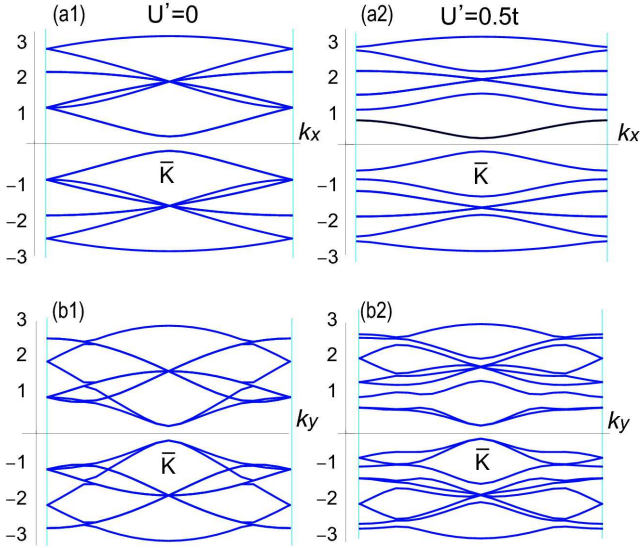}}
\caption{(Color online) Band structure of the ($3\times 3$) superstructure
(a1,a2) along the $k_{x}$-axis and (b1,b2) along the $k_{y}$-axis, where
(a1,b1) for $U^{\prime }=0$ and (a2,b2) for $U^{\prime }=0.5t$. Band
structure (a1,b1) agrees with (a3,b3) in Fig.\protect\ref{FigS2Sa}. The
vertical axis is the energy in unit of $t$. The horizontal axis is momentum $%
k_{x}$ or $k_{y}$. We have set $\protect\lambda _{\text{SO}}=0.2t$ for
illustration. }
\label{FigBand}
\end{figure}

The total Hamiltonian is $H=H_{0}+H_{1}$, where the spin $s_{z}$ is a good
quantum number. Hence, the Hamiltonian is decomposed into the spin sectors
indexed by $s_{z}=\pm 1$. Since the number (\ref{NumbeSI}) of silicon atoms
per unit cell is $18$ in the (3$\times $3) superstructure, we analyze the
18-band tight-binding model. We show numerically calculated band structures
of the superstructure for $U^{\prime }=0$ and $U^{\prime }\neq 0$ in Fig.\ref%
{FigBand}. They have qualitatively very similar behaviors though there are
some quantitative difference.

\begin{figure}[t]
\centerline{\includegraphics[width=0.35\textwidth]{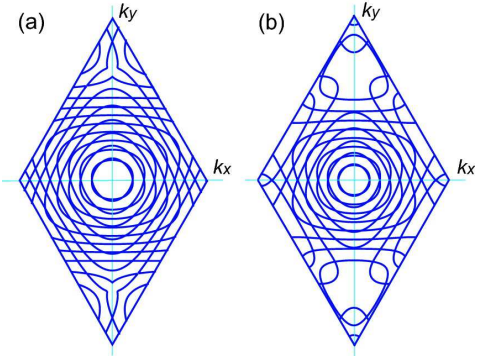}}
\caption{(Color online) Contour plot of two trigonally warped Dirac cones at
the \={K} point (a) for $U^{\prime }=0$ and (b) for $U^{\prime }=0.5t$. They
are originally present at the K and K' points in silicene.}
\label{FigContourJ}
\end{figure}

We also show the contour plot of the band structures (Fig.\ref{FigContourJ}%
). We find two trigonally warped Dirac cones at the center \={K} of the
Brillouin zone. The origin of these two cones is clear: There are Dirac
cones at the K and K' points in silicene, but they are placed on the same
point in the superstructure since the K and K' points are identified. There
appears a new feature because these two Dirac cones now cross each other.
The level crossing will be transformed into the level anticrossing when
Dirac electrons interact among themselves. We show the contour plots of the
Dirac cones in Fig.\ref{FigCoulo}, by introducing an infinitesimal on-site
Coulomb interaction. The two Dirac cones merges and transformed into two
hexagonally warped cones as illustrated in Fig.\ref{FigCoulo}. They are no
longer degenerate, and must be experimentally detectable by angle resolved
photoemission spectroscopy (ARPES).

\subsection{Dirac Theory}

To understand this phenomenon of the hexagonal warping analytically, we
analyze the low energy Dirac theory of the Hamiltonian (\ref{BasicHamil}), 
\begin{eqnarray}
H_{\eta } &=&\hbar v_{\text{F}}\left( k_{x}\tau _{x}+\eta k_{y}\tau
_{y}\right) +\frac{\eta ae^{3\pi \eta i/2}}{4\sqrt{3}}\left( k_{\eta
}^{2}\sigma _{+}+k_{-\eta }^{2}\sigma _{-}\right)  \notag \\
&&-\eta \lambda _{\text{SO}}\sigma _{z}\tau _{z},  \label{DiracTheor}
\end{eqnarray}%
where $\mathbf{\tau }=(\tau _{x},\tau _{y},\tau _{z})$ is the Pauli matrix
of pseudospin associated with the A and B sites. Two Dirac cones are
labelled by $\eta =\pm $, which are originally present at the K point ($\eta
=+$) and the K' point ($\eta =-)$ in silicene. The way of trigonal warping
is represented by the factor $e^{3\pi \eta i/2}$ with the phase being
opposite ($\eta =\pm 1$) between the two types of electrons\cite{Ajiki}. We
assume $U^{\prime }=0$ in the analytical treatment for simplicity.
Nevertheless, we are able to see clearly how two trigonally warped Dirac
cones are transformed into two undegenerate hexagonally warped Dirac cones.

\begin{figure}[t]
\centerline{\includegraphics[width=0.35\textwidth]{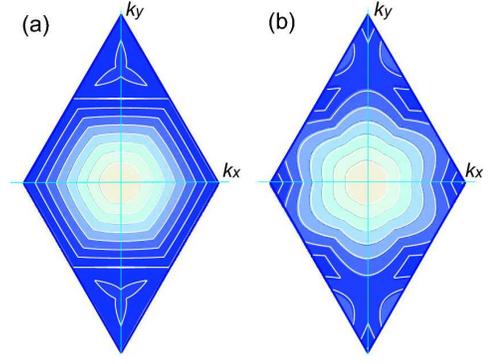}}
\caption{(Color online) Contour plots of two hexagonally warped Dirac cones.
Trigonally warped Dirac cones [Fig.\protect\ref{FigContourJ}] emerge when
two crossing levels are transformed into two anticrossing levels by an
infinitesimal on-site Coulomb interaction. (a) We have taken $U^{\prime }=0$
for simplicity. (b) The Dirac cones are modified slightly when $U^{\prime
}=0.5t$.}
\label{FigCoulo}
\end{figure}

When we introduce the on-site Coulomb interaction, the Hamiltonian $H_{\text{%
K}}$ and $H_{\text{K'}}$ are mixed into%
\begin{equation}
H=\left( 
\begin{array}{cc}
H_{\text{K}} & H_{12} \\ 
H_{21} & H_{\text{K'}}%
\end{array}%
\right) ,
\end{equation}%
where we have chosen the basis $\{c_{A}^{K},c_{B}^{K},c_{A}^{K^{\prime
}},c_{B}^{K^{\prime }}\}$, and%
\begin{equation}
H_{\text{K}}=\left( 
\begin{array}{cc}
-\lambda _{\text{SO}} & \hbar v_{\text{F}}k_{-}+ae^{3\pi i/2}k_{+}^{2} \\ 
\hbar v_{\text{F}}k_{+}+ae^{-3\pi i/2}k_{-}^{2} & \lambda _{\text{SO}}%
\end{array}%
\right) ,
\end{equation}%
\begin{equation}
H_{\text{K'}}=\left( 
\begin{array}{cc}
\lambda _{\text{SO}} & \hbar v_{\text{F}}k_{+}-ae^{-3\pi i/2}k_{-}^{2} \\ 
\hbar v_{\text{F}}k_{-}-ae^{3\pi i/2}k_{+}^{2} & -\lambda _{\text{SO}}%
\end{array}%
\right) .
\end{equation}%
\begin{equation}
H_{12}=H_{21}=\left( 
\begin{array}{cc}
V & 0 \\ 
0 & V%
\end{array}%
\right) ,
\end{equation}%
with $V$ representing the Coulomb energy. Up to the first order in $V$, the
two bands are explicitly given by%
\begin{equation}
\varepsilon _{\pm }^{2}=\left( \hbar v_{\text{F}}k\right) ^{2}+\lambda _{%
\text{SO}}^{2}+a^{2}\pm \sqrt{2}a\hbar v_{\text{F}}k\sqrt{1+\cos 6\theta },
\end{equation}%
where $\theta $ is the azimuthal angle in the momentum space, $k_{x}=k\cos
\theta $. The degeneracy has been resolved. We can check that these two band
structures reproduce the hexagonally warped Dirac cones in Fig.\ref{FigCoulo}
quite well.

\section{Topological Numbers}

The topological quantum numbers\cite{Hasan,Qi} are the Chern number $%
\mathcal{C}$ and the $\mathbb{Z}_{2}$ index. If the spin $s_{z}$ is a good
quantum number, the $\mathbb{Z}_{2}$ index is identical to the spin-Chern
number $\mathcal{C}_{s}$ modulo $2$. They are defined when the state is
gapped and given by 
\begin{equation}
\mathcal{C}=\mathcal{C}_{\uparrow }+\mathcal{C}_{\downarrow },\qquad 
\mathcal{C}_{s}=\frac{1}{2}(\mathcal{C}_{\uparrow }-\mathcal{C}_{\downarrow
}),
\end{equation}%
where $\mathcal{C}_{\uparrow \downarrow }$ is the summation of the Berry
curvature in the momentum space over all occupied states of electrons with $%
s_{z}=\pm 1$. The characteristic feature is that they are unchanged even if
some parameters are continuously switched off in the Hamiltonian provided
that the gap does not close\cite{Prodan09B,Sheng}.

\begin{figure}[t]
\centerline{\includegraphics[width=0.4\textwidth]{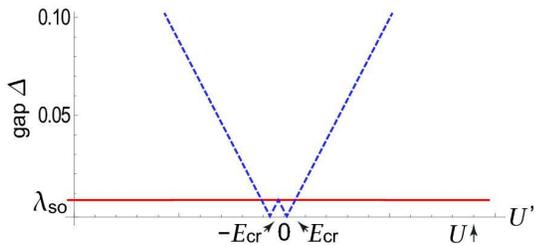}}
\caption{(Color online) Band gap $\Delta $ as a function of $U^{\prime }$.
We show $\Delta $ for the ($3\times 3$) superstructure in the solid curve
and for free-standing silicene in the dashed lines. The band gap closes at $%
U^{\prime }=\pm 2\ell E_{\text{cr}}$ with $E_{\text{cr}}\equiv \protect%
\lambda _{\text{SO}}/\ell $ for the free-standing silicene ($U=0$), but not
closes for the ($3\times 3$) superstructure ($U^{\prime }=U=0.1$eV). The
vertical axis is the energy in unit of $t$. We have set $\protect\lambda _{%
\text{SO}}=0.005t$ for illustration.}
\label{FigGap}
\end{figure}

The spin-Chern number is defined by the integration of the Berry curvature $%
\Omega _{i}(\mathbf{k})$ over the conduction band indexed $i$ in the
Brilloin zone $S$ for each spin sector, 
\begin{equation}
\mathcal{C}_{s_{z}}=\sum_{i}^{N/2}\int_{S}d^{2}k\,\Omega _{i}(\mathbf{k}).
\label{ChernB}
\end{equation}%
Here, the Brilloin zone $S$ is that of the superstructure. The number of
conduction bands is $N/2$ when the system is half filled. By making use of
the band structure (\ref{BrillIdent}) we may equivalently integrate the
Berry curvature $\Omega (\mathbf{k})$ over one conduction band in the
extended Brilloin zone $S_{\text{ex}}$, which is the Brilloin zone of
silicene,%
\begin{equation}
\mathcal{C}_{s_{z}}=\int_{S_{\text{ex}}}d^{2}k\,\Omega (\mathbf{k}).
\end{equation}%
Recall that $N/2$ conduction bands in the Brilloin zone $S$ are merged into
one conduction band in the extended Brilloin zone $S$, as illustrated in Fig.%
\ref{FigS2Sa}.

It is well known that silicene is a SQH insulator characterized by the
topological numbers,%
\begin{equation}
\mathcal{C}=0,\qquad \mathcal{C}_{s}=1.  \label{SiTopNum}
\end{equation}%
It is an important question whether a superstructure is also a topological
insulator. We investigate this problem in the following three ways. All of
them produce the same result on this problem.

\subsection{Adiabatic Analysis}

We take the example of the ($3\times 3$) superstructure by investigating our
Hamiltonian system $H=H_{0}+H_{1}$. We have explicitly calculated the band
gap of the ($3\times 3$) superstructure by taking the potential term (\ref%
{PotenU}), whose result is given in Fig.\ref{FigGap}. It is surprising that
the band gap never closes as a function of $U^{\prime }=U+2\ell E_{z}$. Let
us first choose the external electric field $E_{z}$ to cancel the internal
chemical potential difference $U$, namely, $U^{\prime }=U+2\ell E_{z}=0$.
Then, the Hamiltonian of the system is reduced to that of silicene, and
hence the topological numbers are given by (\ref{SiTopNum}). We then
continuously change $E_{z}$ till $E_{z}=0$. As we have seen, the gap is kept
open during this process. Hence, the topological quantum numbers are kept
unchanged during this continuous process. Namely the ($3\times 3$)
superstructure is topologically connected to silicene. We conclude that the
superstructure is a QSH insulator.

\begin{figure}[t]
\centerline{\includegraphics[width=0.5\textwidth]{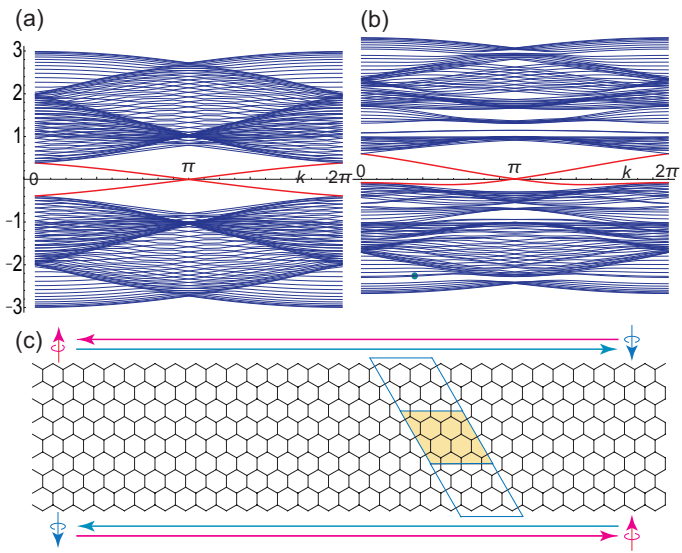}}
\caption{(Color online) Band structure of a zigzag nanorbibbon made of the ($%
3\times 3$) superstructure (a) for $U^{\prime }=0$ and (b) for $U^{\prime
}=0.5t$. The width contains 8 unit cells of the superstructure. The red
lines indicate the doubly degenerate edge modes with up and down spins. The
vertical axis is the energy in unit of $t$. The horizontal axis is the
momentum $k$. The points $k=0$ and $k=2\protect\pi $ are identified, and
give the \={K} point, while $k=\protect\pi $ is the $\Gamma $ point. We have
set $\protect\lambda _{\text{SO}}=0.2t$ for illustration. Each red line
connects the \={K} point in the conduction and valence bands after winding
once a circle of the whole Brillouin zone. (c) Illustration of a nanoribon
with the width containing 3 unit cells of the superstructure. The red and
blue arrows indicate the degenerate gapless edge modes with up and down
spins.}
\label{FigRibbon}
\end{figure}

When we apply external electric field $E_{z}$, silicene undergoes a
topological phase transition\cite{EzawaNJP}. However, the ($3\times 3$)
superstructure does not since the gap does not close. We shall see the
reason why this is the case in subsection \ref{SubsecDirac}.

\subsection{Edge States Analysis}

The bulk-edge correspondence\cite{Hasan,Qi} is well known to characterize
the topological insulators. Namely, a topological insulator is characterized
by the appearance of gapless modes on edges. The reason why gapless modes
appear in the edge of a topological insulator is understood as follows. The
topological insulator has a nontrivial topological number, the $\mathbb{Z}%
_{2}$ index\cite{KaneMele}, which is defined only for a gapped state. When a
topological insulator has an edge beyond which the region has the trivial $%
\mathbb{Z}_{2}$ index, the band must close and yield gapless modes in the
interface. Otherwise the $\mathbb{Z}_{2}$ index cannot change its value
across the interface. 

We have numerically calculated the band structure of a zigzag nanoribbon
geometry in Fig.\ref{FigRibbon}(a,b), where the width contains 8 unit cells
of the superstructure. Two zero-energy edge modes (indicated by red lines)
are clearly seen: Each line represents the two-fold degenerate helical modes
propagating one edge as indicated in Fig.\ref{FigRibbon}(c). The band
structure of a nanoribbon with $U^{\prime }=0$ is just the one obtained by
folding that of a silicene nanoribbon as in Fig.\ref{FigRibbon}(a). It is
modified for $U^{\prime }\neq 0$, but the modification is only slight even
when we take a quite large value for $U^{\prime }$, as shown in Fig.\ref%
{FigRibbon}(b). The essential feature is that the edge state, connecting the
conduction and valence bands at the same \={K} point, is topologically
protected since it winds a circle of the whole Brillouin zone as in Fig.\ref%
{FigRibbon}. Namely, as far as the band is open, though the band structure
is modified by changing $U^{\prime }$, the gapless edge modes connecting the
conduction and valence bands continue to exist. We remark that each zero
mode originally connects the K and K' points in silicene and is
topologically protected. This property is not lost even when they are
identified as the \={K} point. Consequently, the superstructure is a QSH
insulator.

\subsection{Dirac Theory Analysis}

\label{SubsecDirac}

We have found that there exists no topological phase transition in the ($%
3\times 3$) superstructure by the change of the external electric field $%
E_{z}$. We present an analytic argument by calculating the topological
numbers based on the effective Dirac theory. This analysis is valid for a
generic superstructure. We assume that the buckled structure consists of two
sublattices. Indeed, there is a report\cite{Enriquez} that this is the case
for all structures grown on Ag(1,1,1). Let the numbers of A sites and B
sites be $N_{A}$ and $N_{B}$, respectively, in the higher sublattice per
unit cell which contains $N$ silicon atoms. It yields a contribution to the
Dirac mass $m_{\text{D}}$ within the mean-field approximation,%
\begin{equation}
m_{\text{D}}=\frac{N_{A}-N_{B}}{N}(U+2\ell E_{z})+\eta s_{z}\lambda _{\text{%
SO}}.  \label{MassFormu}
\end{equation}%
It has been shown\cite{EzawaEPJB,Hasan,Qi} that the topological charges are
determined by the sign of the Dirac mass $m_{\text{D}}$.

In the ($3\times 3$) superstructure, since we have $N_{A}=N_{B}=3$ and $N=18$%
, the superstructure is a QSH insulator. Furthermore, since it does not
depend on $E_{z}$, there occurs no phase transitions by the change of $E_{z}$%
. This gives an analytic reasoning why the gap does not close in Fig.\ref%
{FigGap}.

The condition that the phase transition occurs \cite{EzawaEPJB} is that $m_{%
\text{D}}=0$, which determines the critical field, 
\begin{equation}
E_{\text{cr}}=-\frac{1}{2\ell }\left( \eta s_{z}\lambda _{\text{SO}}\frac{N}{%
N_{A}-N_{B}}+U\right) .
\end{equation}%
In the case of free-standing silicene, we have $U=0$, $N=2$, $N_{A}=1$, $%
N_{B}=0$, and hence 
\begin{equation}
E_{\text{cr}}=\lambda _{\text{SO}}/\ell ,
\end{equation}%
as agrees with the previous result\cite{EzawaNJP}. In general it is
necessary that $N_{A}\neq N_{B}$ for a topological phase transition to occur
in a superstructure described by the potential term (\ref{PotenU}).

\section{Discussions}

\label{SecDiscu}

\begin{figure}[t]
\centerline{\includegraphics[width=0.35\textwidth]{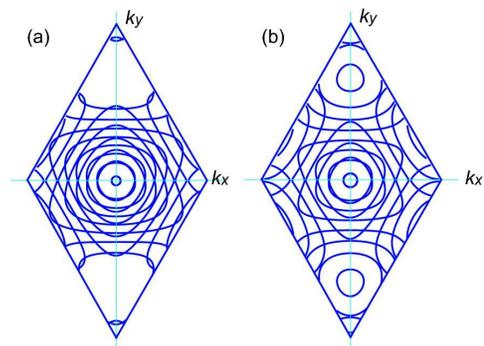}}
\caption{(Color online) Cotour plot of Dirac cones with structural
deformation by transfer-energy difference for (a) $U^{\prime }=0$ and (b) $%
U^{\prime }=0.5t$. We set $t_{\text{LL}}=1.5t$ and $t_{\text{LH}}=t$. By
comparing these with Fig.\protect\ref{FigContourJ}, the modification of the
band structure is found to be quite small.}
\label{FigDeformed}
\end{figure}

In conclusion, we have investigated the topological properties of a
superstructure made from the silicene honeycomb lattice by a periodic
potential term $H_{1}$. The superstructure is a QSH insulator if it is
topologically connected to silicene. The concept of the topological
connectedness may be formulated as follows. We consider the Hamiltonian,%
\begin{equation}
H(\lambda )=H_{0}+\lambda H_{1}.
\end{equation}%
Provided the gap is open in the system $H(\lambda )$ for any values of $%
\lambda $ ($0\leq \lambda \leq 1$) as $\lambda $ is continuously changed
from $\lambda =1$ to $\lambda =0$, the two systems $H(1)$ and $H(0)$ are
topologically connected. For the instance of the potential (\ref{PotenU}) we
have used the electric field $E_{z}$ instead of $\lambda $ as such a
continuous parameter to show the topological connectedness.

We have investigated the problem if the topological phase transition occurs
as in silicene\cite{EzawaNJP}. We have also derived the condition (\ref%
{CondiOneK}) that the K and K' points become identical in superstructures.
In this case, there occurs an intriguing phenomenon that two Dirac cones
coexist at a single point (\={K} point). They exhibit hexagonal warping in
the presence of the Coulomb interaction, which must be observed by ARPES.

We have carried out a numerical analysis and confirm these observations by
taking an instance of the ($3\times 3$) buckled superstructure together with
the potential term $H_{1}$ given by (\ref{PotenU}). In so doing we have
assumed that all transfer energys are the same. In principle, there may be
two different transfer energyies, $t_{\text{LL}}$ and $t_{\text{LH}}$, where 
$t_{\text{LL}}$ is the transfer energy between two lower sites (indicated as
blue balls) and $t_{\text{LH}}$ is the transfer energy between lower and
higher sites (indicated as blue and red balls) in Fig.\ref{FigUnit},
although there is no report that there is a significant difference between
them. Note that there is no $t_{\text{HH}}$ term as seen in Fig.\ref{FigUnit}%
. We have calculated the band structure with different transfer energy in
Fig.\ref{FigDeformed}. It is clearly seen that there is no qualitative
difference even if we assume very different transfer energies such as $t_{%
\text{LL}}=1.5t_{\text{LH}}=1.5t$. A possible transfer-energy difference
does not cause any qualitative change of the band structure. In particular,
the gap does not close as a function of $E_{z}$. We conclude that such a
transfer-energy difference does not destroy the topological properties of
the superstructure.

Silicene superstructures are grown on the Ag substrate. The most common one
is the ($3\times 3$) superstructure on the ($4\times 4$) Ag-structure. There
are several possible superstructures obtained in this way. Here we give some
correspondences between them\cite{Jamgo,Enriquez}:%
\begin{equation*}
\begin{tabular}{|c|c|c|c|c|c|}
\hline
Si-superstructure & Ag-structure & p & q & ratio & Refs. \\ \hline
$3\times 3$ & $4\times 4$ & $3$ & $0$ & $1.33$ & \cite{GLayPRL,Kawai} \\ 
\hline
$2\times 2$ & $\sqrt{7}\times \sqrt{7}$ & $2$ & $0$ & $1.32$ &  \\ \hline
$\sqrt{7}\times \sqrt{7}$ & $\sqrt{13}\times \sqrt{13}$ & $2$ & $1$ & $1.36$
& \cite{Kawai} \\ \hline
$2\sqrt{3}\times 2\sqrt{3}$ & $\sqrt{21}\times \sqrt{21}$ & $2$ & $2$ & $%
1.32 $ &  \\ \hline
$\sqrt{7}\times \sqrt{7}$ & $2\sqrt{3}\times 2\sqrt{3}$ & $2$ & $1$ & $1.31$
& \cite{Jamgo} \\ \hline
\end{tabular}%
\end{equation*}%
It is interesting that the ratios of the Ag-structure and Si-superstructures
are around $1.3$. There would be other ways to construct superstructures by
introducing periodic potentials to silicene. Our analysis of topological
properties in superstructure is applicable to any of them.

I am very much grateful to G.L. Lay, S. Hasegawa and N. Nagaosa for many
fruitful discussions on the subject. This work was supported in part by
Grants-in-Aid for Scientific Research from the Ministry of Education,
Science, Sports and Culture No. 22740196.


\begin{thebibliography}{99}
\bibitem{NoteA} It is customary for experimentalists to call the
superstructure of silicene by the corresponding structure of the Ag
substrate. Thus the ($3\times 3$) superstructure of silicene is referred to
as the ($4\times 4$) structure of silver, where $3\times 3$ silicon atoms
are on top of $4\times 4$ silver atoms.
\end{thebibliography}
\end{document}